PAPER • OPEN ACCESS

# Phase control of the fractional conductance of silicon nanosandwich-structures

View the article online for updates and enhancements.

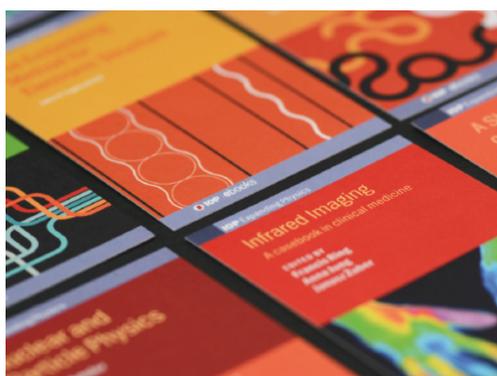

# Phase control of the fractional conductance of silicon nanosandwich-structures

**N I Rul'[1,2], N T Bagraev[1,2], L E Klyachkin[2] and A M Malyarenko[2]**

[1] Peter the Great St. Petersburg Polytechnic University, St. Petersburg, 195251, Russia
[2] Ioffe Institute, St. Petersburg 194021, Russia

E-mail: rul.nickolai@mail.ru

**Abstract**. We present the experimental data of the electric features of the silicon nanosandwichstructures obtained by silicon planar technology in the frameworks of the Hall geometry that represent the ultra-shallow silicon quantum well of 2 nm wide that are confined by delta-barrier heavily doped with boron, which create the edge channels used as the phase controllers of electric signals. The formation of the negative-U dipole boron centers, which appear to confine the edge channels, results in the effective mass dropping and corresponding reduction of the electron-electron interaction thereby giving rise to the macroscopic quantum phenomena at high temperatures up to room temperature. The phase control of the longitudinal conductance is observed by changing either the magnitude of the source-drain current or the voltage applied to the external gate of the silicon nanosandwiches within the quantum Faraday effect.

## 1. Introduction

Nanostructures which are based on silicon planar nanosandwiches prepared in the frameworks of the Hall geometry are of particular interest as various devices such as phase controllers of electric signals. These silicon nanosandwiches represent the ultra-shallow silicon quantum wells of 2 nm wide that are confined by delta-barriers heavily doped with boron, which are prepared by silicon planar technology [1] (figure 1). The high concentration of boron inside the delta-barrier, $N(B) = 5 \cdot 10^{21}$ cm$^{-3}$, on the one hand and the small sheet density in the quantum well, $n_{2D} = 3 \cdot 10^{13}$ m$^{-2}$, on the other hand appear to give rise to the formation of the topological edge channels [2]. It should be noted that such high concentration of boron in the delta-barriers confining the edge channels of the silicon nanosandwich results in the formation of the negative-U trigonal dipole boron centers, with the significant reduction of the electron-electron interaction [3]. This important achievement that is accompanied by dropping an effective mass of carriers to such values as $m^* \approx 2 \cdot 10^{-5} m_e$ causes the observation of quantum interference in the edge channel at the high temperatures up to room temperature [4]. Thus, the goal of this work is to study the high temperature macroscopic quantum phenomena such as phase characteristics of conductance, which are controlled by changing either the magnitude of the source-drain current or the voltage applied to the external gate of the silicon quantum nanosandwiches obtained in the frameworks of the Hall geometry.

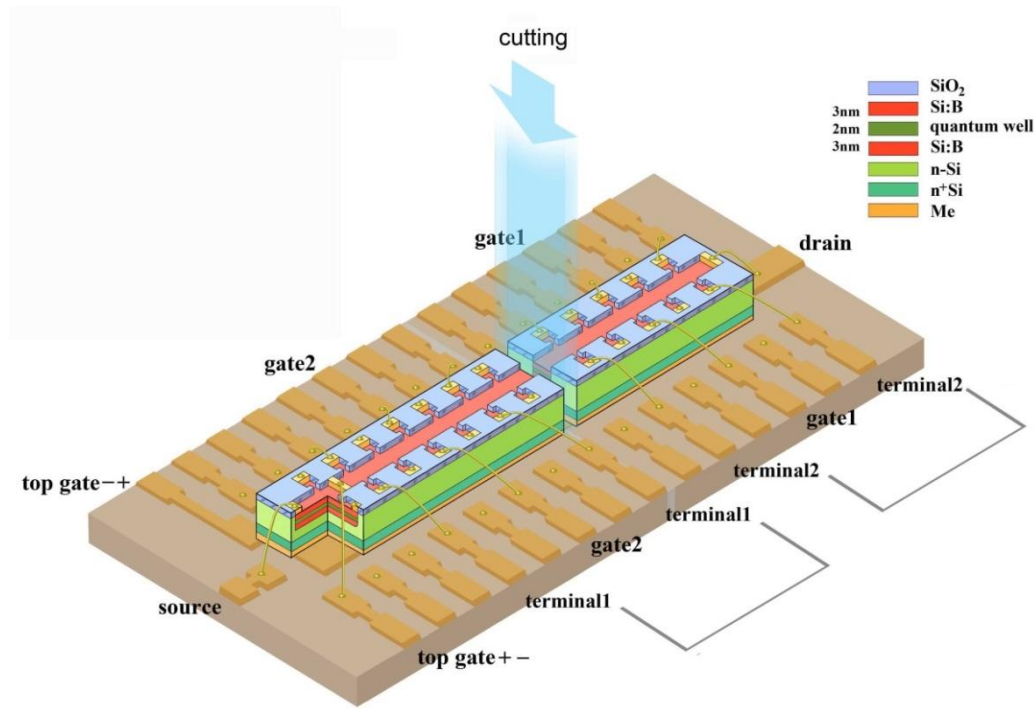

**Figure 1.** Silicon nanosandwich represents the ultra-shallow quantum well of 2 nm wide that is confined by delta-barriers heavily doped with boron, which are prepared by silicon planar technology. This device structure is of particular interest both as the phase invertor and modulator of the quantum conductance characteristics.

## 2. Fractional conductance quantization

The current-voltage characteristics of the edge channels of silicon nanosandwiches show the fractional quantization of the longitudinal conductance, which is dependent on the magnitude of the stabilized source-drain current, $I_{ds}$, as the quantizing parameter on the one hand and the lateral gate voltage applied to the Hall contacts as the control parameter on the other hand (figure 2). The fact is that the presence of the stabilized source-drain current results in the appearance of the magnetic field induced by the system of non-interacting carriers in the absence of the electron-electron interaction and an external magnetic field. Within the frameworks of the quantum harmonic oscillator model that is applicable to each single carrier in the edge channel, induced magnetic field is proportional to the magnitude of the source-drain current, $B = \beta I_{ds}$. Therefore the subsequence of the energy levels in the edge channel becomes more complex, because the Landau levels created by induced magnetic field are added to the dimensional quantization levels, with the energy difference proportional to the source-drain current value. Thus, it can be argued that the oscillations of the longitudinal conductance as a function of the source-drain current values seem to represent an analogue of the Shubnikov – de Haas effect. It is worth noting that the coefficient of proportionality $\beta$ is not always constant and appears to be dependent on the source-drain current. In this case, it is well known that the maximum value of any kinetic coefficient will be observed when the Fermi level passes through the middle of the corresponding Landau level. For this reason, the peak position of the conductance maximum value has to correspond to the condition

$$E_F = E_i + \hbar e \beta I_{ds}(n + 0.5)/m^*, \tag{1}$$

where $E_i$ denotes the dimensional quantization level, at which the Fermi level determined by the sheet density is in the good agreement with the energy position of the current induced Landau levels in the absence of an external magnetic field.

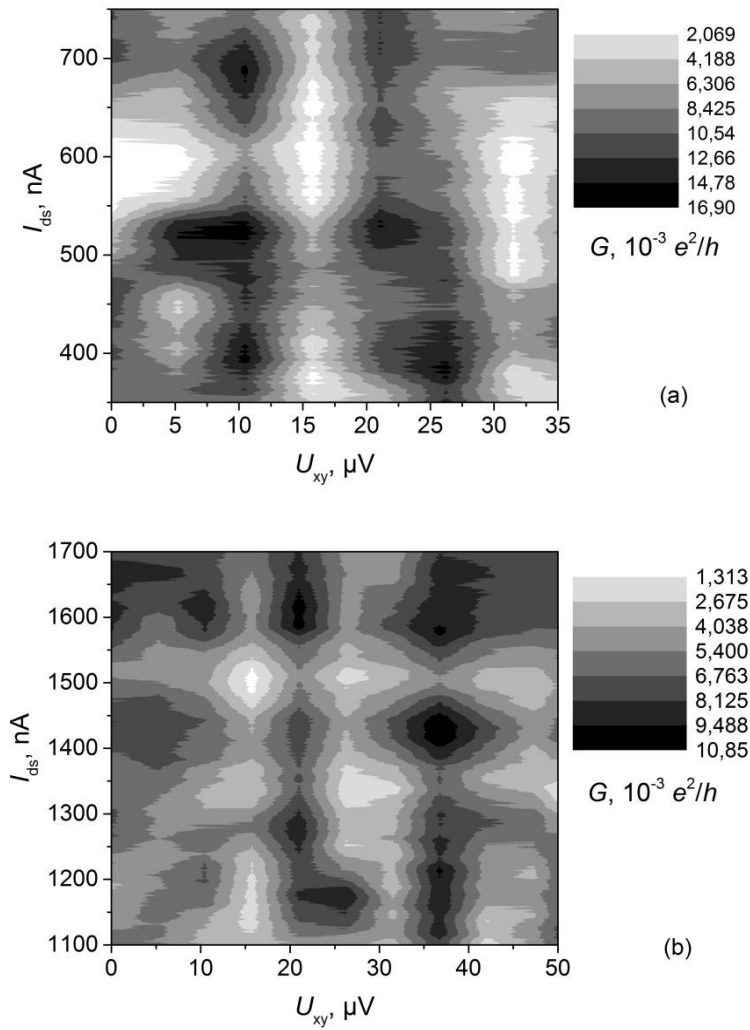

**Figure 2.** The fractional quantized longitudinal conductance of the edge channels of silicon nanosandwiches show both dependence on the magnitude of the stabilized source-drain current as the quantizing parameter and the transverse gate voltage applied to the Hall contacts as the control parameter.

Besides, the magnetic field induced by the source-drain current results in the appearance of the diamagnetic current $I_{\text{dia}}$, which is not involved in the charge transfer, but in case of the voltage applied to the lateral gates is capable to effect on the magnitude of the longitudinal conductance within the quantum Faraday effect. Owing to the above noticed, the significant reduction of the electron-electron interaction appears to provide the process of the single magnetic flux quanta capture on the system of non-interacting carriers in the edge channel within the frameworks of the Laughlin model [2, 5, 6]. Thus, by varying the value of the voltage applied to the lateral gate, the phase shift in the Aharonov – Bohm effect,

$$G = G_0(I_{\text{ds}}, U_{\text{xy}}) \cdot cos^2(\pi \Phi'/2\Phi_0), \qquad (2)$$

can be created. The total magnetic flux $\Phi' = \Phi + \Delta\Phi$, which determines the conductance phase, depends both on the magnitude of the source-drain current, $\Phi = \alpha I_{\text{ds}}$, and on the value of the lateral gate voltage, $\Delta\Phi = -eU_{xy}/I_{dia}$, with the linear dependence of the diamagnetic current on the source-drain current

$$I_{dia} = I_{dia}(I_{ds}) = \gamma I_{ds}. \quad (3)$$

The legitimacy of the formula used is determined not only by its agreement with experimental data obtained ($\alpha = 4.32 \cdot 10^{-5}$ Wb/A, $\gamma \approx 6 \cdot 10^{-4}$), but also by the maximum value of the longitudinal conductance, which is about three percent of the conductance quantum value $G_0 = e^2/h$, which is evidence of the weak coupling regime and of weak localization in the edge channel of silicon nanosandwich (figure 3).

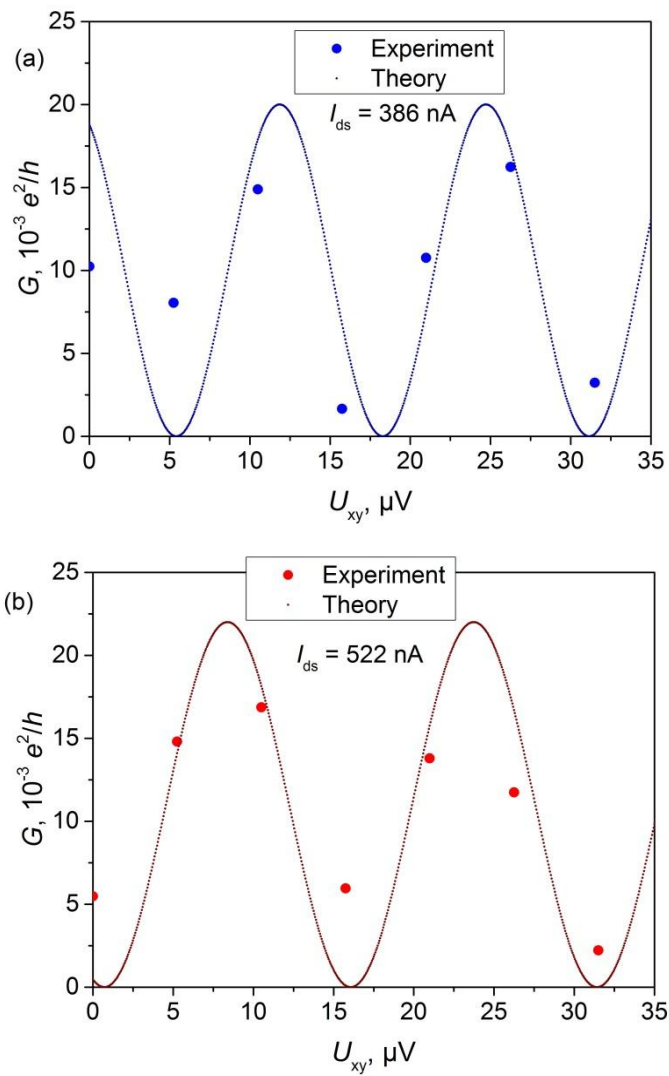
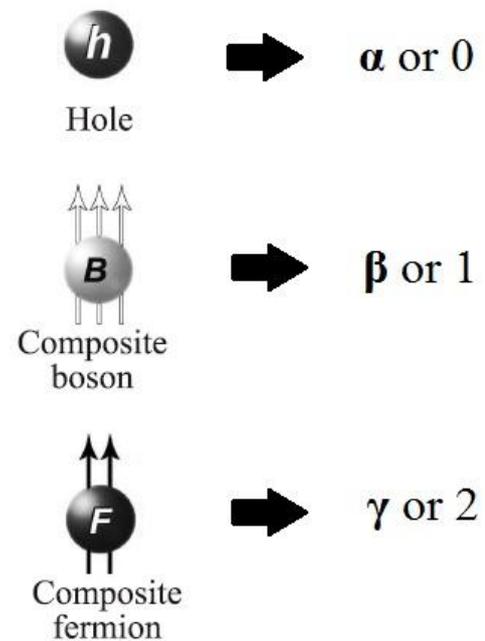

**Figure 3.** The phase shift in the magnitude of the longitudinal conductance, which is similar to the phase shift in the Aharonov-Bohm effect.

**Figure 4.** The alphabet based on the composite particles – bosons and fermions – which allows the classic logical element based on quantum processes.

### 3. Logical elements based on the silicon nanosandwich

The process of the single magnetic flux quanta capture on the system on non-interacting carriers in the edge channel gives an opportunity for logical elements construction. Such logical elements represent the classic logical elements based on the quantum processes caused by each single carrier, which is able to store information. Thus, any message of any length can be created with using the alphabet based on composite particles (figure 4), which are dependent on both the magnitude of the source-drain current and lateral gate voltage and determined by the number of captured induced magnetic flux quanta per number of the carriers. It should be noted that the formation of fractional composite particles enhance the possibility of this alphabet to transfer the information flows, with regulation of the entropy in the presence of the negative-U dressing for the edge channels.

### 4. Conclusions

In summary, the phase characteristics of the longitudinal conductance of the edge channels of silicon quantum nanosandwich determine the legitimacy of their interpretation within the quantum Faraday effect, with the ability of phase control by either the source-drain current or the external gale voltage changing. Finally, the classic logical elements based on the composite particles alphabet are shown to be experimentally realized owing to the negative-U entropy control.